\documentclass[11pt]{article}
\usepackage{authblk}
\usepackage[]{inputenc}
\usepackage{amsfonts}
\usepackage{latexsym,amsmath}
\usepackage{amssymb,array}
\makeatletter \oddsidemargin 0in \evensidemargin 0in \textwidth 16cm
 \RequirePackage[dvips]{graphicx} \textheight 18cm
\begin{document}
\newcommand{\wt}{\widetilde}
\newtheorem{thm}{Theorem}[section]
\newtheorem{pro}{Proposition}[section]
\newtheorem{remark}{Remark}[section]
\newtheorem{ce}{Counterexample}[section]
\newtheorem{cor}{Corollary}[section]
\newtheorem{d1}{Definition}[section]
\newtheorem{ex}{Example}[section]
\newtheorem{lem}{Lemma}[section]
\numberwithin{equation}{section}
\title{\Large{Coherent systems with dependent and identically distributed components: A study of relative ageing based on cumulative hazard and cumulative reversed hazard rate functions}}
\author[a]{Nil Kamal Hazra\footnote{Corresponding author, email: nilkamal@iitj.ac.in}}
\author[b]{ Neeraj Misra}
\affil[a]{Department of Mathematics, Indian Institute of Technology Jodhpur, Karwar-342037, Rajasthan, India}
\affil[b]{Department of Mathematics and Statistics, Indian Institute of Technology Kanpur, Kanpur 208016, India}
\date{}
\maketitle
\begin{abstract}
The relative ageing is an important notion which is useful to measure how a system ages relative to another one. Among all existing stochastic orders, there are two important orders describing the relative ageing of two systems, namely, ageing faster orders in the cumulative hazard and the cumulative reversed hazard rate functions. In this paper, we give some sufficient conditions under which one coherent system ages faster than another one with respect to aforementioned stochastic orders. Further, we show that the proposed sufficient conditions are satisfied for $k$-out-of-$n$ systems. Moreover, some numerical examples are given to illustrate the developed results.
\end{abstract}
{\bf Keywords:} Coherent system, dual distortion/domination function, $k$-out-of-$n$ system, stochastic orders
\\{\bf 2010 Mathematics Subject Classification:} Primary 90B25
\\\hspace*{3.2 in}Secondary 60E15; 60K10
\section{Introduction and Preliminaries}\label{se1}
Stochastic ageing is one of the important concepts in reliability theory and survival analysis. It broadly describes the pattern in which a system ages/degrades with time. In the literature, three different types of ageing notions were introduced, namely, positive ageing, negative ageing and no ageing. Positive ageing means that the residual lifetime of a system decreases with time, stochastically, whereas the negative ageing describes the completely opposite phenomenon. On the other hand, no ageing means that the system does not age with time. Based on these three ageing notions, different types of ageing classes (namely, IFR, DFR, IFRA, DFRA, ILR, DLR, to name a few) have been proposed in the literature in order to study different ageing characteristics of a system. For more discussion on this topic, the reader may go through Barlow and Proschan~\cite{bp}, and Lai and Xie~\cite{lx}. Apart from these ageing concepts, relative ageing is an another useful notion which has tremendous applications in the study of system reliability. Relative ageing measures how a sytem improves/deteriorates over time with respect to another system.
\\\hspace*{0.2 in}There are many real-life scenarios where we deal with more than one system of the same type (for example, television sets of different companies, CPUs produced by different brands, etc.). In such cases, we often come across the following question: How to measure if one system ages faster compared to others as time progresses? An effective solution to this question is provided through the notion of relative ageing. Another aspect of the relative ageing is found to be useful when we deal with the phenomenon of crossing hazards/mean residual lifetimes problems. These kinds of problems occur in many real-life situations. For example, Pocock et al.~\cite{pgk} noticed the crossing hazards phenomenon when they were dealing with the survival data on the effects of two different treatments to breast cancer patients. Further, Champlin et al.~\cite{cmeg} also
reported some instances where the supremacy of a treatment over another treatment stays for a while. Thus, the above discussion suggests that the increasing/decreasing hazard ratio models could be considered as a reasonable choice in many real-life situations. Indeed, Kalashnikov and Rachev~\cite{kr} introduced a realtive ageing notion (called ageing faster order in the hazard rate) based on monotonicity of the ratio of two hazard rate functions. In the same spirit, Sengupta and Deshpande~\cite{sd} proposed another notion which is defined on monotonicity of the ratio of two cumulative hazard rate functions. Later, Di Crescenzo~\cite{d} developed two relative ageing notions based on monotonicities of the ratios of two reversed hazard and two cumulative reversed hazard functions, whereas Finkelstein~\cite{f6} proposed a notion based on the concept of mean residual lifetime functions. Furthermore, Hazra and Nanda~\cite{hn0} introduced some relative ageing notions based on some generalized functions.  
\\\hspace*{0.2 in} In what follows, we discuss coherent systems. A system is said to be coherent if each of its components is relevant, and its structure function is monotonically nondecreasing with respect to each argument (see Barlow and Prochan~\cite{bp} for the definition). The so called $k$-out-of-$n$ system is a special case of the coherent system. A $k$-out-of-$n$ system is a system, formed by $n$ components, that functions as long as at least $k$ of its $n$ components function. Further, the series ($n$-out-of-$n$) and the parallel ($1$-out-of-$n$) systems are special cases of the $k$-out-of-$n$ system. Moreover, the lifetime of a $k$-out-of-$n$ system is the same as $(n-k+1)$-th order statistic of the lifetimes of $n$ components. Thus, the study of a $k$-out-of-$n$ system and the study of an $(n-k+1)$-th order statistic (of nonnegative random variables) are basically the same.
\\\hspace*{0.2 in}It is meaningful to study coherent systems because these are the skeletons of most of the real-life systems. Thus, the study of different reliability properties of coherent systems is considered as one of the important problems in reliability theory. The problem, which we focus here, is the study of relative ageing of coherent systems. There exists a vast literurate on various stochastic comparisons of $k$-out-of-$n$ systems (in particular, series and parallel systems) with independent components (see, for instance, Pledger and Proschan~\cite{pp}, Proschan and Sethuraman~\cite{ps}, Balakrishnan and Zhao~\cite{bz}, and the references therein). Further, the same problem for coherent systems was considered in Esary and Proschan~\cite{ep}, Nanda et al.~\cite{njs}, Kochar et al.~\cite{kms}, Belzunce et al.~\cite{bfrr}, Navarro and Rubio~\cite{nr}, Navarro et al.~\cite{nass1, nass3, npd}, Samaniego and Navarro~\cite{sn}, Amini-Seresht et al.~\cite{azb}, to name a few. However, all of these studies were done in terms of traditional stochastic orders, namely, usual stochastic order, hazard rate order, reversed hazard rate order, likelihood ratio order, etc. There are a few articles where ageing faster orders were considered to compare two coherent systems. The study of $k$-out-of-$n$ systems using ageing faster orders (in the hazard and the reversed hazard rates) was done by Misra and Francis~\cite{mf}, Li and Li~\cite{ll6}, and Ding and Zhang~\cite{dz}, whereas a similar study for coherent systems was done by Ding et al.~\cite{dfz}, and Hazra and Misra~\cite{hm}. Recently, Misra and Francis~\cite{mf2} studied $k$-out-of-$n$ systems with independent components in terms of ageing faster orders (in the cumulative hazard and the cumulative reversed hazard rates). However, this problem is not yet considered for coherent systems with dependent components. Thus, the goal of this paper is to study coherent systems with dependent components in terms of ageing faster orders in the cumulative hazard and the cumulative reversed hazard rates.
\\\hspace*{0.2 in}Now, we will introduce some notation and acronyms that will be used throughout the paper. For a random variable $W$ (with absolutely continuous distribution function), we denote its probability density function (pdf) by $f_W(\cdot)$, the cumulative distribution function (cdf) by $F_W(\cdot)$, the survival/reliability
 function by $\bar F_W(\cdot)\equiv 1-F_W(\cdot)$, the hazard rate function by $r_W(\cdot)\equiv f_W(\cdot)/\bar F_W(\cdot)$, the reversed hazard rate function by $\tilde r_W(\cdot)\equiv f_W(\cdot)/F_W(\cdot)$, the cumulative hazard rate function by $\Delta_W(\cdot)\equiv -\ln \bar F_W(\cdot)$, and the cumulative reversed hazard rate function by $\wt\Delta_W(\cdot)\equiv -\ln F_W(\cdot)$. Further, for any two random variables $U$ and $V$, we write $U\stackrel{d}=V$ to mean that $U$ and $V$ are identically distributed. Furthermore, we denote the set of real numbers by $\mathbb{R}$. By $a\stackrel{\rm sgn}=b$ we mean that $a$ and $b$ have the same sign, whereas $a\stackrel{\rm def.}=b$ means that $b$ is defined as $a$. We use two acronyms i.i.d and d.i.d. for `independent and identically distributed' and `dependent and identically distributed', respectively.
 \\\hspace*{0.2 in}Let $\tau\left(\mbox{\boldmath$X$}\right)$ be a random variable representing the lifetime of a coherent system made up of $n$ components with the lifetime vector $\mbox{\boldmath$X$}=(X_1,X_2,\dots,X_n)$, where $X_i$'s are d.i.d. For the sake of notational simplicity, let us assume that $X_i\stackrel{d}=X$, $i=1,2,\dots,n$, for some non-negative random variable $X$. Then the joint reliability function of $\mbox{\boldmath$X$}$ can be written as 
 \begin{eqnarray*}
\bar F_{\mathbf{X}}(x_1,x_2,\dots,x_n)&=&P\left(X_1>x_1,X_2>x_2,\dots,X_n>x_n\right)
\\&=&K\left(\bar F_{X}(x_1),\bar F_{X}(x_2),\dots,\bar F_{X}(x_n)\right),\quad x_i\in \mathbb{R},\;i=1,2,\dots,n,
\end{eqnarray*}
where $K(\cdot,\cdot,\dots,\cdot)$ is a survival copula. It is worthwhile to mention here that this representation indeed follows from Sklar's Theorem (see Nelsen~\cite{n}). The survival copula describes the dependency structure among components of a system. In the literature, different types of survival copulas have been proposed, for example, Farlie-Gumbel-Morgenstern (FGM) copula, Archimedean copula with different generators,
Clayton-Oakes (CO) copula, etc. (see Nelsen~\cite{n} for more discussion on this topic). Below we give a useful lemma that represents a relation between the reliability function of a system and the reliability functions of its corresponding components.
 \begin{lem}[Navarro et al.~\cite{nass1}]\label{II}
 Let $\tau\left(\mbox{\boldmath$X$}\right)$ be the lifetime of a coherent system formed by $n$ d.i.d. components with the lifetime vector $\mbox{\boldmath$X$}=(X_1,X_2,\dots,X_n)$. Then the reliability function of $\tau\left(\mbox{\boldmath$X$}\right)$ can be written as
 $$\bar F_{\tau\left(\mbox{\boldmath$X$}\right)}(x)=P(\tau\left(\mbox{\boldmath$X$}\right)>x)=h\left(\bar F_{X}(x)\right),\quad x\in \mathbb{R},$$
 where $h(\cdot):[0,1]\rightarrow [0,1]$, called the domination (or dual distortion) function, depends on the structure function $\phi(\cdot)$ (see Barlow and Proschan~\cite{bp} for definition) and on the survival copula $K$ of $X_1,X_2,\dots,X_n$. Furthermore, $h(\cdot)$ is an increasing continuous function in $[0,1]$ such that $h(0)=0$ and $h(1)=1$. $\hfill\Box$
\end{lem}
  \hspace*{0.2 in} In what follows, we discuss some preliminary concepts that will be helpful in better understanding of the paper. We begin with the notion of stochastic orders which are effectively used to compare the lifetimes of two systems. In the literature, different types of stochastic orders have been developed to compare two random variables describing two different random phenomena in different branches of mathematics and statistics. The detailed discussion on this topic could be found in the monographs written by Shaked and Shanthikumar~\cite{ss}, and Belzunce et al.~\cite{bmr}. Below we give the definitions of stochastic orders that are used in this paper.
\begin{d1}\label{de1}
Let $X$ and $Y$ be two random variables with absolutely continuous distribution functions supported on $[l_X,u_X]$ and $[l_Y,u_Y]$, respectively,
where $u_X$ and $u_Y$ may be $\infty$, and $l_X$ and $l_Y$ may be $-\infty$, with the convention that $[-\infty, a]\stackrel{\rm def.}=(-\infty,a]$ for all $a\in\mathbb{R}$, and $[b,\infty]\stackrel{\rm def.}=[b,\infty)$ for all $b\in\mathbb{R}$.
Then, $X$ is said to be smaller than $Y$ in the
\begin{enumerate}
\item [$(a)$] hazard rate (hr) order, denoted as $X\leq_{hr}Y$, if $${\bar F_Y(x)}/{\bar F_X(x)}\;\text{ is increasing in } x \in (-\infty,max[u_X,u_Y]);$$
\item [$(b)$] reversed hazard rate (rh) order, denoted as $X\leq_{rh}Y$, if $$ {F_Y(x)}/{ F_X(x)}\;\text{ is increasing  in } x \in(min[l_X,l_Y],\infty);$$
 \item [$(c)$] usual stochastic (st) order, denoted as $X\leq_{st}Y$, if $$\bar F_X(x)\leq \bar F_Y(x) \text{ for all, }x\in \mathbb{R};$$
\end{enumerate}
 here, for any positive real constant $a$, $a/0\stackrel{\rm def.}=\infty$.
\end{d1}
It is worthwhile to mention here that
$$ X\leq_{hr [rh]}Y \implies X\leq_{st}Y.$$
\hspace*{0.2 in}Apart from above discussed stochastic orders, there are two more sets of stochastic orders which are frequently used to study the relative ageings of two systems. One of them is the set of transform orders (for example, convex transform order, super-additive order, star-shaped order, DMRL order, $s$-IFR order, $s$-IFRA orders, etc.) which are defined based on different ageing concepts, namely, increasing failure rate, increasing failure rate in average, new better than used, etc. (see, for instance, Barlow and Proschan~\cite{bp}, Deshpande and Kochar~\cite{dk}, Bartoszewicz~\cite{b}, Kochar and Wiens~\cite{kw}, and Nanda et al.~\cite{nhag}). The other one is the set of ageing faster orders that describe whether a system is ageing faster than another one in terms of the hazard rate, the cumulative hazard rate, the reversed hazard rate, the cumulative reversed hazard rate, etc. (see, Kalashnikov and Rachev~\cite{kr}, Sengupta and Deshpande~\cite{sd}, Di Crescenzo~\cite{d}, Finkelstein~(\cite{f6}, \cite{f8}), Razaei et al.~\cite{rgi}, Hazra and Nanda~\cite{hn0}, Misra et al.~\cite{mfn}, Kayid et al.~\cite{kiz}, Misra and Francis~\cite{mf2}, and the references therein). For the sake of completeness in our presentation, we give the definitions of following stochastic orders.
\begin{d1}
Let $X$ and $Y$ be two random variables with absolutely continuous distribution functions supported on $[0,\infty)$.
Then $X$ is said to be ageing faster than $Y$ in the
\begin{enumerate}
\item [$(a)$] hazard rate, denoted as $X\underset{ c}{\prec} Y$, if
$$r_X(x)/r_Y(x)\text{ is increasing in }x\in[0,\infty);$$
\item [$(b)$] reversed hazard rate, denoted as $X\underset{ b}{\prec} Y$, if
$$\tilde r_X(x)/\tilde r_Y(x)\text{ is decreasing in }x\in[0,\infty);$$
\item [$(c)$] cumulative hazard rate, denoted as $X\underset{ c^*}{\prec} Y$, if
$$\Delta_X(x)/\Delta_Y(x)\text{ is increasing in }x\in[0,\infty);$$
\item [$(d)$] cumulative reversed hazard rate, denoted as $X\underset{ b^*}{\prec} Y$, if
$$\wt \Delta_X(x)/\wt \Delta_Y(x)\text{ is decreasing in }x\in[0,\infty).$$
\end{enumerate}
\end{d1}
It is known that
$$ X\underset{ c}{\prec} Y \implies X\underset{ c^*}{\prec} Y,\text{ and }X\underset{ b}{\prec} Y \implies X\underset{ b^*}{\prec} Y.$$
\hspace*{0.2 in}Relative ageing of $k$-out-of-$n$ systems comprising of components having i.i.d. lifetimes has been studied by many researchers (for references see the paragraph preceding Definition 1.2). Under the i.i.d. set-up, Misra and Francis~\cite{mf2} established the following results:
\begin{itemize}
\item [$(a)$] $ X\underset{ c^*}{\prec} Y$ and $Y\leq_{st}X$ $\implies$ $X_{k:n}\underset{ c^*}{\prec}Y_{k:n},\quad k=1,2,\dots,n$,
\item [$(b)$] $ X\underset{ b^*}{\prec} Y$ and $X\leq_{st}Y$ $\implies$ $X_{k:n}\underset{ b^*}{\prec}Y_{k:n},\quad k=1,2,\dots,n$,
\end{itemize}
where $X_{k:n}$ (resp. $Y_{k:n}$) denotes the $k$-th order statistic based on i.i.d. observations $X_1,X_2,\dots,X_n$ (resp. $Y_1,Y_2,\dots,Y_n$). The purpose of this paper is to derive more general results that accommodate general coherent systems comprising of components having d.i.d. lifetimes.
\\\hspace*{0.2 in}In what follows, we give the definitions of TP$_2$ and RR$_2$ functions which are used in the proofs of the main results (cf. Karlin~\cite{k}).
\begin{d1}
 Let $\mathcal{X}$ and $\mathcal{Y}$ be two linearly ordered sets. Then, a real-valued and nonnegative function $\kappa(\cdot,\cdot)$, defined on $\mathcal{X}\times\mathcal{Y}$,
 is said to be TP$_2$ (resp. RR$_2$) if
 $$\kappa(x_1,y_1)\kappa(x_2,y_2)\geq(\text{resp. }\leq)\;\kappa(x_1,y_2)\kappa(x_2,y_1),$$
  for all $x_1\leq x_2$ and $y_1\leq y_2$.$\hfill \Box$
\end{d1}
\hspace*{0.2 in} Throughout the paper, the words increasing, decreasing, positive and negative are not used in the strict sense.  All random variables considered in this paper are assumed to have absolutely continuous distribution functions supported on $[0,\infty)$. Further, we use bold symbols to represent vectors (e.g., $\mbox{\boldmath$b$}=(b_1,b_2,\dots,b_p)$, $b_i\in \mathbb{R}$, $i=1,2,\dots,p$). For positive integers $k,l,m$ and $n$ ($1\leq k\leq n$, $1\leq l\leq m$), we use $\tau_{k|n}$ and $\tau_{l|m}$ to represent the lifetimes of $k$-out-of-$n$ and $l$-out-of-$m$ systems, respectively. 
\\\hspace*{0.2 in}The rest of the paper is organized as follows. In Section~\ref{s2}, we discuss some important lemmas that are used in proving the main results. In Section~\ref{s3}, we discuss main results. Here, we provide some sufficient conditions under which one coherent system performs better than another one with respect to ageing faster orders in terms of the cumulative hazard and the cumulative reversed hazard rates. Further, we show that the proposed results hold for the well known $k$-out-of-$n$ systems thereby generalizing the results of Misra and Francis~\cite{mf2} to the situations where we have general coherent systems comprising of components with d.i.d. lifetimes. We also provide some examples to illustrate the applications of the proposed results. 
\\\hspace*{0.2 in}All proofs of lemmas and theorems, wherever given, are deferred to the Appendix.  
 \section{Useful Lemmas}~\label{s2}
 In this section we present some lemmas which will be used in proving the main results. The first lemma is borrowed from Karlin~(\cite{k}, Theorem 11.2, pp. 324-325), and Hazra and Nanda~(\cite{hnx}, Lemma 3.5), whereas next two lemmas were obtained in Hazra and Misra~\cite{hm}.
\begin{lem}\label{l0}
 Let $\kappa(x,y)>0$, defined on $\mathcal{X}\times \mathcal{Y}$, be RR$_2$ (resp. TP$_2$), where $\mathcal{X}$ and $\mathcal{Y}$ are subsets of the real line. 
 Assume that a function $f(\cdot,\cdot)$ defined on $\mathcal{X}\times\mathcal{Y}$ is such that
 \begin{itemize}
  \item [$(i)$] for each $x\in \mathcal{X}$, $f(x,y)$ changes sign at most once, and if the change of sign does occur, it is from positive to negative, as $y$ traverses $\mathcal{Y}$;
  \item [$(ii)$] for each $y \in \mathcal{Y}$, $f(x,y)$ is increasing (resp. decreasing) in $x\in \mathcal{X}$;
  \item [$(iii)$] $\omega(x)=\int\limits_{\mathcal{Y}}\kappa(x,y)f(x,y)d\mu(y)$ exists absolutely and defines a continuous function of $x$, where $\mu$ is a sigma-finite measure. 
 \end{itemize}
Then $\omega(x)$ changes sign at most once, and if the change of sign does occur, it is from negative (resp. positive) to positive (resp. negative), as $x$ traverses $\mathcal{X}$.
\end{lem}
\begin{lem}\label{l3}
Let $h_{k|n}(\cdot)$ and $h_{l|m}(\cdot)$ be the reliability functions of the $k$-out-of-$n$ and the $l$-out-of-$m$ systems with i.i.d. components, respectively, where $1\leq k\leq n$ and $1\leq l\leq m$. Further, let $H_{k|n}(p)=ph_{k|n}'(p)/h_{k|n}(p)$ and $H_{l|m}(p)=ph_{l|m}'(p)/h_{l|m}(p)$, for all $p\in(0,1)$. Then the following results hold.
\begin{itemize}
  \item [$(i)$] ${H_{k|n}(p)}/{H_{l|m}(p)}$ is decreasing in $p\in(0,1)$, for all $k\leq l$ and $m-l\leq n-k$; 
  \item [$(ii)$] $(1-p){H_{k|n}'(p)}/{H_{k|n}(p)}$ is negative and decreasing in $p\in (0,1)$.
  \end{itemize}
\end{lem}
\begin{lem}\label{l4}
Let $h_{k|n}(\cdot)$ and $h_{l|m}(\cdot)$ be the reliability functions of the $k$-out-of-$n$ and the $l$-out-of-$m$ systems with i.i.d. components, respectively, where $1\leq k\leq n$ and $1\leq l\leq m$. Further, let $R_{k|n}(p)=(1-p)h_{k|n}'(p)/(1-h_{k|n}(p))$ and $R_{l|m}(p)=(1-p)h_{l|m}'(p)/(1-h_{l|m}(p))$, for all $p\in(0,1)$. Then the following results hold.
\begin{itemize}
  \item [$(i)$] ${R_{k|n}(p)}/{R_{l|m}(p)}$ is increasing in $p\in(0,1)$, for all $l\leq k$ and $n-k\leq m-l$; 
  \item [$(ii)$] $p{R_{k|n}'(p)}/{R_{k|n}(p)}$ is positive and decreasing in $p\in (0,1)$.
  \end{itemize}
\end{lem}
\section{Main Results}\label{s3}
Let $\tau_1\left(\mbox{\boldmath$X$}\right)$ and $\tau_2\left(\mbox{\boldmath$Y$}\right)$ be the lifetimes of two coherent systems formed by two different sets of d.i.d. components with the lifetime vectors $\mbox{\boldmath$X$}=(X_1,X_2,\dots,X_{n})$ and $\mbox{\boldmath$Y$}=(Y_1,Y_2,\dots,Y_{m})$, respectively. For the sake of simplicity of notation, let us assume that $X_i\stackrel{\rm d}=X$, $i=1,2,\dots,n,$ and $Y_j\stackrel{\rm d}=Y$, $j=1,2,\dots,m,$ for some non-negative random variables $X$ and $Y$. Further, let $h_1(\cdot)$ and $h_2(\cdot)$ be the domination functions of $\tau_1\left(\mbox{\boldmath$X$}\right)$ and $\tau_2\left(\mbox{\boldmath$Y$}\right)$, respectively.
In what follows, we use the following notation:
   $$H_i(p)=\frac{ph_i'(p)}{h_i(p)}\text{ and }R_i(p)=\frac{(1-p)h_i'(p)}{1-h_i(p)},\quad p\in(0,1),\;i=1,2.$$ 
   Note that when $m=n$ and the two coherent systems have the same structure function as well as the same survival copula (i.e., the two coherent systems are identical), then $\tau_1(\cdot)=\tau_2(\cdot)=\tau(\cdot)$ (say), $h_1(\cdot)=h_2(\cdot)=h(\cdot)$ (say), $H_1(\cdot)=H_2(\cdot)=H(\cdot)$ (say) and $R_1(\cdot)=R_2(\cdot)=R(\cdot)$ (say). 
\subsection{Relative ageing based on cumulative hazard functions}
In this subsection, we compare two coherent systems in terms of ageing faster order in the cumulative hazard rate.
\\\hspace*{0.2 in} In the following proposition we consider a coherent system and provide sufficient conditions under which use of components with lifetimes vector $\mbox{\boldmath$X$}$ in the coherent system makes it age faster than if components with lifetimes vector $\mbox{\boldmath$Y$}$ are used in the coherent system. 
 \begin{pro}\label{t1}
Assume that $m=n$ and that the two coheret systems have the same structure function and the same survival copula. Further suppose that the following conditions hold:
\begin{itemize}
\item [$(i)$] $(1-p){H'(p)}/{H(p)}$ is negative and decreasing in $p\in (0,1)$;
\item [$(ii)$] $X\underset{ c^*}{\prec} Y$ and $Y\leq_{st} X $.
\end{itemize}
Then $\tau\left(\mbox{\boldmath$X$}\right)\underset{ c^*}{\prec}\tau\left(\mbox{\boldmath$Y$}\right)$.$\hfill\Box$
\end{pro}
\hspace*{0.3 in}In the following theorem we compare two coherent systems with different domination functions, and generalize the result of Proposition~\ref{t1}. Note that the domination functions of two coherent systems differ when they have either different structure functions or different dependency relations among components, or both. The proof of the theorem follows from Proposition~3.1 of Hazra and Misra~\cite{hm}, Proposition~\ref{t1} and by using the facts that $\underset{ c^*}{\prec}$ is transitive and $\underset{ c}{\prec}\implies \underset{ c^*}{\prec}$.
\begin{thm}\label{t11}
Suppose that $\{(i),(ii),(iv)\}$ or $\{(i),(iii),(iv)\}$ holds:
\begin{itemize}
\item [$(i)$] ${H_1(p)}/{H_2(p)}$ is decreasing in $p\in(0,1)$;
\item [$(ii)$] $(1-p){H_1'(p)}/{H_1(p)}$ is negative and decreasing in $p\in (0,1)$;
\item [$(iii)$] $(1-p){H_2'(p)}/{H_2(p)}$ is negative and decreasing in $p\in (0,1)$;
\item [$(iv)$] $X\underset{ c^*}{\prec} Y$ and $Y\leq_{st} X $.
\end{itemize}
Then $\tau_1\left(\mbox{\boldmath$X$}\right)\underset{ c^*}{\prec}\tau_2\left(\mbox{\boldmath$Y$}\right)$.
\end{thm}
\begin{remark}
It is worth mentioning here that the results stated in Proposition~\ref{t1} and Theorem~\ref{t11} also follow from Theorem 3.1 of Hazra and Misra~\cite{hm} under the same set of conditions, as in Theorem~\ref{t11}, except that there condition $(iv)$ of Theorem~\ref{t11} is replaced by the condition $X\underset{ c}{\prec} Y$ and $Y\leq_{rh} X $. Note that this assumption is much stronger as compared to $(iv)$ because $\underset{ c}{\prec}\implies \underset{ c^*}{\prec}$ and $ \leq_{rh} \implies \leq_{st}$, and the reverse implications are not necessarily true. The methodology used in the proofs of this paper differs from that in Hazra and Misra~\cite{hm}. Thus, it is meaningful to study the results given in Proposition~\ref{t1} and Theorem~\ref{t11}.$\hfill\Box$
\end{remark}
\hspace*{0.2 in} The following corollary follows from Theorem~\ref{t11} and Lemma~\ref{l3}. It shows that the result stated in Theorem~\ref{t11} indeed holds for $k$-out-of-$n$ systems with i.i.d. components. Further, it is to be noted that Theorem 3.1 of Misra and Francis~\cite{mf2} is a particular case of this corollary ($k=l$ and $m=n$).
  \begin{cor}\label{c1}
  Suppose that the $X_i$'s are i.i.d., and that the $Y_j$'s are i.i.d.
  If $X\underset{ c^*}{\prec} Y$ and $Y\leq_{st} X $, then
  ${\tau_{k|n}\left(\mbox{\boldmath$X$}\right)}\underset{ c^*}{\prec}$ ${\tau_{l|m}\left(\mbox{\boldmath$Y$}\right)}$, for all $k\leq l$ and $m-l\leq n-k$.$\hfill\Box$
  \end{cor}
\hspace*{0.2 in} Below we give an example to demonstrate the applications of Proposition~\ref{t1} and Theorem~\ref{t11}.
\begin{ex}\label{e1}
Let us consider two coherent systems $\tau_1(\mbox{\boldmath$X$})=\min\{X_1,\max\{X_2,X_3\}\}$ and $\tau_2(\mbox{\boldmath$Y$})=\min\{Y_1,Y_2,Y_3\}$ formed by two different sets of components with the lifetime vectors $\mbox{\boldmath$X$}=(X_1,X_2,X_3)$ and $\mbox{\boldmath$Y$}=(Y_1,Y_2,Y_3)$, respectively. Assume that $X_i\stackrel{d}=X$ and $Y_i\stackrel{d}=Y$, for all $i=1,2,3,$ and for some non-negative random variables $X$ and $Y$. Further, let $\bar F_X(x)=\exp\{-\alpha_1\left(x+\beta x^2\right)\}$, $x>0$, and $\bar F_Y(x)=\exp\{-\alpha_2\left(x+\beta x^2\right)\}$, $x>0$, where $0<\alpha_1\leq \alpha_2$ and $\beta>0$. Then, it can easily be verified that $X\underset{ c^*}{\prec} Y$ and $Y\leq_{st} X $. Assume that $Y_1,Y_2$ and $Y_3$ are independent whereas $X_1,X_2$ and $X_3$ are dependent and their joint survival function is described by the FGM copula 
  $$K(p_1,p_2,p_3)=p_1p_2p_3(1+\theta (1-p_1)(1-p_2)(1-p_3)),$$ 
  where $p_i\in(0,1)$, $i=1,2,3$, and $\theta\in [-1,1]$.  
Then the domination functions of $\tau_1(\mbox{\boldmath$X$})$ and $\tau_2(\mbox{\boldmath$Y$})$ are given by
 \begin{eqnarray*}
  h_1(p)=2p^2-p^3-\theta p^3 (1-p)^3 \text{ and }h_2(p)=p^3,\quad 0<p<1,
  \end{eqnarray*}
 respectively. From the above expressions, we get
  \begin{eqnarray*}
  \frac{(1-p)H_2'(p)}{H_2(p)}=0 \text{ and }\frac{H_1(p)}{H_2(p)}=\frac{4-3(1+\theta)p+12\theta p^2-15\theta p^3+6\theta p^4}{6-3(1+\theta)p+9\theta p^2-9\theta p^3+3\theta p^4}, \quad 0<p<1.
  \end{eqnarray*}
 Note that condition ($i$) of Theorem~\ref{t1} holds (see Example 3.1 of Navarro et al.~\cite{nass1}) whereas
condition ($iii$) of Theorem~\ref{t1} is vaciously satisfied. Hence $\tau_1\left(\mbox{\boldmath$X$}\right)\underset{ c^*}{\prec}\tau_2\left(\mbox{\boldmath$Y$}\right)$ follows from Theorem~\ref{t1}.
\end{ex}
  \begin{remark}
It is to be noted that the condition $Y\leq_{st} X $ given in Proposition~\ref{t1}, Theorem~\ref{t11} and Corollary~\ref{c1} cannot be dropped as illustrated in Example 3.2 of Misra and Francis~\cite{mf2}.
\end{remark}
  \subsection{Relative ageing based on cumulative reversed hazard functions}
  In this subsection, we compare two coherent systems with respect to the ageing faster order in the cumulative reversed hazard rate.
  \\\hspace*{0.2 in} In the following proposition we consider the situation where $m=n$ and the two coherent systems have the same structure function as well as the same survival copula, i.e., the two coherent systems are identical. We derive sufficient conditions under which use of components having lifetimes vector $\mbox{\boldmath$X$}$ makes the system age faster than if the components with lifetimes vector $\mbox{\boldmath$Y$}$ are used. 
   \begin{pro}\label{t2}
Assume that $m=n$ and that the two coherent systems have the same structure function and the same survival copula. Further suppose that the following conditions hold:
\begin{itemize}
\item [$(i)$] $p{R'(p)}/{R(p)}$ is positive and decreasing in $p\in (0,1)$;
\item [$(ii)$] $X\underset{ b^*}{\prec} Y$ and $X\leq_{st} Y $.
\end{itemize}
Then $\tau\left(\mbox{\boldmath$X$}\right)\underset{ b^*}{\prec}\tau\left(\mbox{\boldmath$Y$}\right)$.$\hfill\Box$
\end{pro}
\hspace*{0.3 in}The following theorem is an extension of the above result. Here we compare two coherent systems with different domination functions. The proof follows from Proposition 3.2 of Hazra and Misra~\cite{hm} and Proposition~\ref{t2} and by using the facts that $\underset{ b^*}{\prec}$ is transitive and $\underset{ b}{\prec}\implies \underset{ b^*}{\prec}$.
\begin{thm}\label{t22}
Suppose that $\{(i),(ii),(iv)\}$ or $\{(i),(iii),(iv)\}$ holds:
\begin{itemize}
\item [$(i)$] ${R_1(p)}/{R_2(p)}$ is increasing in $p\in(0,1)$;
\item [$(ii)$] $p{R_1'(p)}/{R_1(p)}$ is positive and decreasing in $p\in (0,1)$;
\item [$(iii)$] $p{R_2'(p)}/{R_2(p)}$ is positive and decreasing in $p\in (0,1)$;
\item [$(iv)$] $X\underset{ b^*}{\prec} Y$ and $X\leq_{st} Y$.
\end{itemize}
Then $\tau_1\left(\mbox{\boldmath$X$}\right)\underset{ b^*}{\prec}\tau_2\left(\mbox{\boldmath$Y$}\right)$.
\end{thm}
\begin{remark}
It is to be mentioned here that the results given in Proposition~\ref{t2} and Theorem~\ref{t22} also follow from Theorem~3.2 of Hazra and Misra~\cite{hm} under the same set of conditions, as in Theorem~\ref{t22}, except that there condition $(iv)$ above is replaced by $X\underset{ b}{\prec} Y$ and $X\leq_{hr} Y $. Note that this assumption is much stronger as compared to $(iv)$ because $\underset{ b}{\prec}\implies \underset{ b^*}{\prec}$ and $ \leq_{hr} \implies \leq_{st}$, and the reverse implications are not necessarily true. Thus, the study of the results given in Proposition~\ref{t2} and Theorem \ref{t22} is worthy of investigation.$\hfill\Box$
\end{remark}
\hspace*{0.2 in} The following corollary immediately follows from Theorem~\ref{t22} and Lemma~\ref{l4}. Note that Theorem 4.1 of Misra and Francis~\cite{mf2} is a particular case of this corollary ($k=l$ and $m=n$).
 \begin{cor}\label{c2}
  Suppose that the $X_i$'s are i.i.d., and that the $Y_j$'s are i.i.d.
  If $X\underset{ b^*}{\prec} Y$ and $X\leq_{st} Y $, then 
${\tau_{k|n}\left(\mbox{\boldmath$X$}\right)}\underset{ b^*}{\prec}$ ${\tau_{l|m}\left(\mbox{\boldmath$Y$}\right)}$, for all $l\leq k$ and $ n-k\leq m-l$.$\hfill\Box$
  \end{cor}
  \hspace*{0.2 in}The following example illustrates the results given in Proposition~\ref{t2} and Theorem \ref{t22}.
  \begin{ex}
Let us consider two coherent systems $\tau_1(\mbox{\boldmath$X$})=\min\{X_1,X_2,\dots,X_m\}$ and $\tau_2(\mbox{\boldmath$Y$})=\min\{Y_1,Y_2,\dots,Y_n\}$ formed by two different sets of components with the lifetime vectors $\mbox{\boldmath$X$}=(X_1,X_2,$ $\dots,X_m)$ and $\mbox{\boldmath$Y$}=(Y_1,Y_2,\dots,Y_n)$, respectively, where $n\leq m$. Assume that $X_i\stackrel{d}=X$, $i=1,2,\dots,m$, and $Y_j\stackrel{d}=Y$, $j=1,2,\dots,n$, for some non-negative random variables $X$ and $Y$. Further, let $\bar F_X(x)=\exp\{-3x\}$, $x>0$, and $\bar F_Y(x)=\exp\{-2x\}$, $x>0$.  Then, it can easily be verified that $X\underset{ b}{\prec} Y$, and hence $X\underset{ b^*}{\prec} Y$. Also, note that $X\leq_{st} Y $. Let us assume that $\mbox{\boldmath$X$}=(X_1,X_2,\dots,X_m)$ has the Gumbel-Hougard survival copula given by
  $$K(p_1,p_2,\dots,p_{m})=\exp\left\{-\left(\sum\limits_{i=1}^{m}(-\ln p_i)^\theta\right)^{1/\theta}\right\},\quad \theta\geq 1\text{ and }0<p_i<1,\text{ for }i=1,2,\dots,m,$$
  and $\mbox{\boldmath$Y$}=(Y_1,Y_2,\dots,Y_n)$ has the Gumbel-Hougard survival copula given by
  $$K(p_1,p_2,\dots,p_{n})=\exp\left\{-\left(\sum\limits_{i=1}^{n}(-\ln p_i)^\theta\right)^{1/\theta}\right\},\quad \theta\geq 1\text{ and }0<p_i<1,\text{ for }i=1,2,\dots,n.$$
  Then the domination functions of ${\tau_1\left(\mbox{\boldmath$X$}\right)}$ and $\tau_2(\mbox{\boldmath$Y$})$ are given by
  $h_1(p)=p^{a}$ and $h_2(p)=p^{b}$, respectively,
  where $a=m^{1/\theta}$ and $b=n^{1/\theta}$, and $1\leq b\leq a<\infty$. Thus we have
   $$R_1(p)=\frac{a(1-p)p^{a-1}}{1-p^a}\text{ and }R_2(p)=\frac{b(1-p)p^{b-1}}{1-p^b},\quad 0<p<1\text{ and }1\leq b\leq a<\infty.$$
  Let $d=a-b$ $(0\leq d<a)$. Then, from the above expressions, we get
    $$l(p)\stackrel{\rm def.}=\frac{R_1(p)}{R_2(p)}=\frac{a}{b}\left(1-\frac{1-p^d}{1-p^a}\right),\quad 0<p<1, \;1\leq b\leq a<\infty \text{ and } 0\leq d<a,$$
    and
    $$\frac{pR_1'(p)}{R_1(p)}=\frac{a-1-ap+p^a}{1-p-p^a+p^{a+1}},\quad 0<p<1\text{ and }1\leq a<\infty.$$
    Now, 
    $$l'(p)=\frac{a}{b}\left[\frac{d(1-p^a)p^{d-1}-a(1-p^d)p^{a-1}}{(1-p^a)^2}\right],\quad 0<p<1, \;1\leq b\leq a<\infty \text{ and } 0\leq d<a.$$
    Since, for $0\leq d<a$,
    $$\frac{1-p^a}{1-p^d}\geq \frac{a}{d}p^{a-d},\quad0<p<1,$$
    it follows that $l'(p)\geq 0$, for all $0<p<1$, and hence $R_1(p)/R_2(p)$ is increasing in $p\in(0,1)$.
     Further, Example 4.1 of Hazra and Misra~\cite{hm} shows that $pR_1'(p)/R_1(p)$ is positive and decreasing in $p\in(0,1)$. Thus, $\tau_1\left(\mbox{\boldmath$X$}\right)\underset{ b^*}{\prec}\tau_2\left(\mbox{\boldmath$Y$}\right)$ follows Theorem~\ref{t22}.
\end{ex}
  \begin{remark}
It is to be noted that the condition $X\leq_{st} Y $ given in Proposition~\ref{t2}, Theorem~\ref{t22} and Corollary~\ref{c2} cannot be relaxed as shown in Example 4.2 of Misra and Francis~\cite{mf2}.
\end{remark}
  \subsection*{Acknowledgments}
\hspace*{0.2 in}The first author sincerely acknowledges the financial support received from IIT Jodhpur, India.
  
 \newpage
 {\bf Appendix}
\\\\{\bf Proof of Proposition~\ref{t1}:} For $x\in(0,\infty)$, we have
\begin{eqnarray*}
\Delta_{\tau\left(\mbox{\boldmath$X$}\right)}(x)&=&\int\limits_0^x r_{\tau\left(\mbox{\boldmath$X$}\right)}(u)du=\int\limits_0^x r_X(u) H\left(\bar F_X(u)\right)du
\\&=&\int\limits_0^x H\left(\bar F_X(u)\right)d\left(-\ln \bar F_X(u)\right)
\\&=&\int\limits_0^{-\ln \bar F_X(x)} H\left(e^{-v}\right)dv.
\end{eqnarray*}
Similarly,
\begin{eqnarray*}
\Delta_{\tau\left(\mbox{\boldmath$Y$}\right)}(x)&=&\int\limits_0^{-\ln \bar F_Y(x)} H\left(e^{-v}\right)dv,\quad x\in(0,\infty).
\end{eqnarray*}
Since $Y\leq_{st} X $, we have $\frac{-\ln \bar F_X(x)}{-\ln \bar F_Y(x)}\leq 1$, for all $x\in(0,\infty)$. Therefore
\allowdisplaybreaks{
\begin{eqnarray*}
\frac{\Delta_{\tau\left(\mbox{\boldmath$X$}\right)}(x)}{\Delta_{\tau\left(\mbox{\boldmath$Y$}\right)}(x)}&=&\frac{\int\limits_0^{-\ln \bar F_X(x)} H\left(e^{-v}\right)dv}{\int\limits_0^{-\ln \bar F_Y(x)} H\left(e^{-v}\right)dv}
\\&=&\frac{\int\limits_0^{\frac{-\ln \bar F_X(x)}{-\ln \bar F_Y(x)}} H\left({\bar F_Y}^z(x)\right)dz}{\int\limits_0^{1} H\left({\bar F_Y}^z(x)\right)dz}
\\&=&\frac{\int\limits_0^{1} H\left({\bar F_Y}^z(x)\right)I\left(z\in \left[0,\frac{-\ln \bar F_X(x)}{-\ln \bar F_Y(x)}\right]\right)dz}{\int\limits_0^{1} H\left({\bar F_Y}^z(x)\right)dz},\quad x\in(0,\infty),
\end{eqnarray*}}
where $I(\cdot)$ denotes the indicator function. Let $\alpha$ be any fixed real number. Consider the following relation
\begin{eqnarray*}
\Delta_{\tau\left(\mbox{\boldmath$X$}\right)}(x)-\alpha\Delta_{\tau\left(\mbox{\boldmath$Y$}\right)}(x)\stackrel{sgn}=\int\limits_0^1 \xi_1(x,z)\eta_1(x,z)dz,
\end{eqnarray*}
where, for $(x,z)\in(0,\infty)\times(0,1),$
\begin{eqnarray*}
\xi_1(x,z)=H\left({\bar F_Y}^z(x)\right)
\end{eqnarray*}
and
\begin{eqnarray*}
\eta_1(x,z)=I\left(z\in \left[0,\frac{-\ln \bar F_X(x)}{-\ln \bar F_Y(x)}\right]\right)-\alpha.
\end{eqnarray*}
We will first show that 
\begin{eqnarray}\label{eq6}
\xi_1(x,z) \text{ is RR}_2\text{ in }(x,z)\in (0,\infty)\times (0,1),
\end{eqnarray}
 i.e., for all $0<z_1<z_2<1$,
\begin{eqnarray*}
\frac{H\left({\bar F_Y}^{z_2}(x)\right)}{H\left({\bar F_Y}^{z_1}(x)\right)}\text{ is decreasing in }x\in(0,\infty).
\end{eqnarray*} 
This is equivalent to proving that
\begin{eqnarray}\label{eq2}
\tilde r_{W_2}(x)\left[(1-\bar F_{W_2}(x))\frac{H'(\bar F_{W_2}(x))}{H(\bar F_{W_2}(x))} \right] \geq \tilde r_{W_1}(x)\left[(1-\bar F_{W_1}(x))\frac{H'(\bar F_{W_1}(x))}{H(\bar F_{W_1}(x))} \right],\;x\in(0,\infty), 
  \end{eqnarray}
where $W_i$ is a random variable with the survival function $\bar F_{W_i}(\cdot)\equiv\bar F^{z_i}_Y(\cdot)$, for $i=1,2$. It can easily be verified that $W_2\leq_{lr}W_1$. Consequently,
 \begin{eqnarray}\label{eq3}
  \tilde r_{W_2}(x)\leq \tilde r_{W_1}(x) \text{ and }\bar F_{W_2}(x)\leq \bar F_{W_1}(x), \text{ for all }x>0.
  \end{eqnarray}
 On using \eqref{eq3} and condition (i), we have
  \begin{eqnarray}\label{eq4}
  0\geq (1-\bar F_{W_2}(x))\frac{H'(\bar F_{W_2}(x))}{H(\bar F_{W_2}(x))}\geq (1-\bar F_{W_1}(x))\frac{H'(\bar F_{W_1}(x))}{H(\bar F_{W_1}(x))},\quad x\in(0,\infty).
  \end{eqnarray}
  On combing \eqref{eq3} and \eqref{eq4}, we get \eqref{eq2}, and hence \eqref{eq6} holds.
 Again, from the condition ($ii$), we have $X\underset{ c^*}{\prec} Y$, i.e.,
 $$\frac{-\ln \bar F_X(x)}{-\ln \bar F_Y(x)} \text{ is increasing in }x\in(0, \infty).$$
This implies that
\begin{eqnarray}\label{eq8}
\eta_1(x,z) \text{ is increasing in }x\in(0,\infty), \text{ for all }z\in(0,1).
\end{eqnarray}
Further, note that
 \begin{eqnarray*}
I\left(z\in \left[0,\frac{-\ln \bar F_X(x)}{-\ln \bar F_Y(x)}\right]\right) \text{ is decreasing in }z\in(0,1), \text{ for all }x\in(0,\infty),
\end{eqnarray*}
which is equivalent to the fact that, for every $x\in(0,\infty)$, $\eta_1(x,z)$ changes sign at most once, and if the change of sign does occur, it is from positive to negative, as $z$ traverses from $0$ to $1$. On using this together with \eqref{eq6} and \eqref{eq8} in Lemma~\ref{l0}, we conclude that $\Delta_{\tau\left(\mbox{\boldmath$X$}\right)}(x)-\alpha\Delta_{\tau\left(\mbox{\boldmath$Y$}\right)}(x)$ changes sign at most once, and if the change of sign does occur, it is from negative to positive, as $x$ traverses from $0$ to $\infty$. Thus, $\Delta_{\tau\left(\mbox{\boldmath$X$}\right)}(x)/\Delta_{\tau\left(\mbox{\boldmath$Y$}\right)}(x)$ is increasing in $x\in(0,\infty),$ and hence the result is proved. $\hfill\Box$
\\\\{\bf Proof of Proposition~\ref{t2}:} For all $x\in(0,\infty)$, we have
\begin{eqnarray*}
\wt \Delta_{\tau\left(\mbox{\boldmath$X$}\right)(x)}&=&\int\limits_x^\infty \tilde r_{\tau\left(\mbox{\boldmath$X$}\right)}(u)du=\int\limits_x^\infty \tilde r_X(u) R\left(\bar F_X(u)\right)du
\\&=&\int\limits_x^\infty R\left(\bar F_X(u)\right)d\left(\ln  F_X(u)\right)
\\&=&\int\limits_0^{-\ln  F_X(x)} R\left(1-e^{-v}\right)dv.
\end{eqnarray*}
Similarly,
\begin{eqnarray*}
\wt \Delta_{\tau\left(\mbox{\boldmath$Y$}\right)}(x)&=&\int\limits_0^{-\ln F_Y(x)} R\left(1-e^{-v}\right)dv, \quad x\in(0,\infty).
\end{eqnarray*}
Since $X\leq_{st} Y $, we have $\frac{-\ln F_X(x)}{-\ln  F_Y(x)}\leq 1$, for all $x\in(0,\infty)$.
Therefore 
\allowdisplaybreaks{
\begin{eqnarray*}
\frac{\wt \Delta_{\tau\left(\mbox{\boldmath$X$}\right)}(x)}{\wt \Delta_{\tau\left(\mbox{\boldmath$Y$}\right)}(x)}&=&\frac{\int\limits_0^{-\ln F_X(x)} R\left(1-e^{-v}\right)dv}{\int\limits_0^{-\ln F_Y(x)} R\left(1-e^{-v}\right)dv}
\\&=&\frac{\int\limits_0^{\frac{-\ln F_X(x)}{-\ln F_Y(x)}} R\left(1- F_Y^z(x)\right)dz}{\int\limits_0^{1} R\left(1-F_Y^z(x)\right)dz}
\\&=&\frac{\int\limits_0^{1} R\left(1- F_Y^z(x)\right)I\left(z\in \left[0,\frac{-\ln  F_X(x)}{-\ln F_Y(x)}\right]\right)dz}{\int\limits_0^{1} R\left(1- F_Y^z(x)\right)dz},\quad x\in(0,\infty),
\end{eqnarray*}}
where $I(\cdot)$ denotes the indicator function. Let $\beta$ be any fixed real number. Consider the following relation
\begin{eqnarray*}
\wt \Delta_{\tau\left(\mbox{\boldmath$X$}\right)}(x)-\beta\wt \Delta_{\tau\left(\mbox{\boldmath$Y$}\right)}(x)\stackrel{sgn}=\int\limits_0^1 \xi_2(x,z)\eta_2(x,z)dz,
\end{eqnarray*}
where
\begin{eqnarray*}
\xi_2(x,z)=R\left(1- F_Y^z(x)\right),\quad(x,z)\in(0,\infty)\times(0,1),
\end{eqnarray*}
and
\begin{eqnarray*}
\eta_2(x,z)=I\left(z\in \left[0,\frac{-\ln  F_X(x)}{-\ln  F_Y(x)}\right]\right)-\beta,\quad(x,z)\in(0,\infty)\times(0,1).
\end{eqnarray*}
At first we will show that
\begin{eqnarray}\label{eq9}
\xi_2(x,z) \text{ is TP}_2\text{ in }(x,z)\in (0,\infty)\times (0,1),
\end{eqnarray}
 i.e., for $0<z_1<z_2<1$,
\begin{eqnarray*}
\frac{R\left(1- F_Y^{z_2}(x)\right)}{R\left(1- F_Y^{z_1}(x)\right)}\text{ is increasing in }x\in(0,\infty).
\end{eqnarray*} 
This is equivalent to establishing that
\begin{eqnarray}\label{eq10}
 r_{U_2}(x)\left[\bar F_{U_2}(x)\frac{R'(\bar F_{U_2}(x))}{R(\bar F_{U_2}(x))} \right] \leq  r_{U_1}(x)\left[\bar F_{U_1}(x)\frac{R'(\bar F_{U_1}(x))}{R(\bar F_{U_1}(x))} \right], \quad x\in(0,\infty),
  \end{eqnarray}
where $U_i$ is a random variable with the survival function $\bar F_{U_i}(\cdot)\equiv 1- F^{z_i}_Y(\cdot)$, for $i=1,2$. It can easily be checked that $U_1\leq_{lr}U_2$, and consequently
 \begin{eqnarray}\label{eq11}
   r_{U_2}(x)\leq r_{U_1}(x) \text{ and }\bar F_{U_1}(x)\leq \bar F_{U_2}(x),\quad x\in(0,\infty).
  \end{eqnarray}
 On using \eqref{eq11} and condition (i), we have
  \begin{eqnarray}\label{eq12}
  0\leq \bar F_{U_2}(x)\frac{R'(\bar F_{U_2}(x))}{R(\bar F_{U_2}(x))}  \leq \bar F_{U_1}(x)\frac{R'(\bar F_{U_1}(x))}{R(\bar F_{U_1}(x))} ,\quad x\in(0,\infty).
  \end{eqnarray}
  On combing \eqref{eq11} and \eqref{eq12}, we get \eqref{eq10}, and hence \eqref{eq9} holds.
 Again, from the condition ($ii$), we have $X\underset{ b^*}{\prec} Y$, i.e.,
 $$\frac{-\ln F_X(x)}{-\ln F_Y(x)} \text{ is decreasing in }x\in(0, \infty),$$
which implies that
\begin{eqnarray}\label{eq14}
\eta_2(x,z) \text{ is decreasing in }x\in(0,\infty), \text{ for all }z\in(0,1).
\end{eqnarray}
Further, note that
 \begin{eqnarray*}
I\left(z\in \left[0,\frac{-\ln F_X(x)}{-\ln F_Y(x)}\right]\right) \text{ is decreasing in }z\in(0,1), \text{ for all }x\in(0,\infty).
\end{eqnarray*}
This implies that, for all $x\in(0,\infty)$, $\eta_2(x,z)$ changes sign at most once, and if the change of sign does occur, it is from positive to negative, as $z$ traverses from $0$ to $1$. On using this together with \eqref{eq9} and \eqref{eq14} in Lemma~\ref{l0}, we conclude that $\wt \Delta_{\tau\left(\mbox{\boldmath$X$}\right)}(x)-\beta \wt \Delta_{\tau\left(\mbox{\boldmath$Y$}\right)}(x)$ changes sign at most once, and if the change of sign does occur, it is from positive to negative, as $x$ traverses from $0$ to $\infty$. Thus, $\wt \Delta_{\tau\left(\mbox{\boldmath$X$}\right)}(x)/\wt \Delta_{\tau\left(\mbox{\boldmath$Y$}\right)}(x)$ is decreasing in $x\in(0,\infty),$ and hence the result is proved. $\hfill\Box$
\end{document}